\documentclass[preprint,12pt]{elsarticle}

\usepackage{amssymb}
\usepackage{amsmath}
\usepackage{amsthm}
\usepackage[hang,flushmargin]{footmisc}

\begin{document}

\begin{frontmatter}



\title{Jumping to male-dominated occupations: A novel way to reduce gender wage gap for Chinese women}


\author{Wei Bai}
\author{Zhongtao Yue}
\author{Tao Zhou\footnote{\footnotesize{Corresponding author at: CompleX Lab, University of Electronic Science and Technology of China, Chengdu 611731, China.\\E-mail address: zhutou@ustc.edu (T. Zhou).}}}

\affiliation{
            organization={CompleX Lab, University of Electronic Science and Technology of China},
            addressline={\\Chengdu 611731, China}, 
            }

\begin{abstract}
Occupational segregation is widely considered as one major reason leading to the gender discrimination in labor market. Using large-scale Chinese resume data of online job seekers, we uncover an interesting phenomenon that occupations with higher proportion of men have smaller gender wage gap measured by the female-male ratio on wage. We further show that the severity of occupational segregation in China is low both overall and regionally, and the inter-occupational discrimination is much smaller than the intra-occupational discrimination. That is to say, Chinese women do not face large barriers when changing their occupations. Accordingly, we suggest Chineses women a new way to narrow the gender wage gap: to join male-dominated occupations. Meanwhile, it is worth noticing that although the gender wage gap is smaller in male-dominated occupations, it does not mean that the gender discrimination is smaller there.
\end{abstract}



\begin{keyword}
Gender wage gap \sep occupational segregation \sep Gender inequality
\end{keyword}

\end{frontmatter}


\section{Introduction}
As one of the 17 Sustainable Development Goals, reducing gender inequality is a major policy concern around the world \cite{world2011world}. The gender wage gap is a prominent part of gender inequality. We use the ratio of females' wage to males' (female-male ratio, $r_{fm}$) as the primary measure of the gender wage gap. Globally, $r_{fm} \approx 0.76$ \cite{leach2015gender}. Zhang \textit{et al.} \cite{zhang2008trends} showed that from 1988 to 2004, $r_{fm}$ in China decreased from 0.863 to 0.762. Although there are fluctuations in some years, it is very clear that the gender wage gap keeps widening over time.

Gender wage gap is often closely related to occupational gender segregation (occupational segregation for short) \cite{gross1968plus}, which refers to the fact that workers in the labor market are assigned to different occupational categories due to gender differences, observed as the concentration of most female labor force in some ``feminine" occupations with low wage and low prestige. Occupational segregation is usually considered as a main way of gender discrimination and a major cause of gender wage gap. In terms of professions, male and female students have different educational processes and outcomes even in the same fields. Although STEM occupations are often high-paying, women in them tend to be concentrated in lower-wageing ones \cite{michelmore2016explaining}. Even when women get the highest-wageing jobs in computer science and engineering, they still earn less than their male counterparts. In different regions of the United States, the wage of female doctors is generally lower than that of male \cite{warner2019gender}. In Brazilian tourism industry, women are valued less than men even when they have the same occupational characteristics \cite{guimaraes2016pay}. In finance, although American male and female MBA graduates earn almost the same at the start of their careers, the ratio of females' logarithmic annual wage to males' has been as low as 0.625 after 10 to 16 years \cite{bertrand2010dynamics}. Gender differentiation has also appeared in different sectors in China: the gender wage gap within the political sector is gradually disappearing, while outside is widening \cite{huang_education}. Yang \textit{et al.} \cite{yang2018height} notice that Chinese male job seekers have remarkably higher salary expectation than females. In the American public sector, occupational segregation is still the main cause of wage disparity \cite{mandel2014gender}.

The impact of occupational segregation on gender wage gap also varies by country and generation. Todd \textit{et al.} \cite{todd2012gender} found that occupational segregation increased in Australia from 1995 to 2011. However, Busch \cite{busch2020gender} found that women's earnings in male-dominated occupations increased in Germany between 1992 and 2015. By using data from the European Structure of Earnings Survey (2010), Boll \textit{et al.} \cite{boll2017eu} showed that sectoral isolation has 0\%-15\% explanatory power in salary decomposition. Laine \cite{laine2008segregation} analyzed the impact of different types of occupational segregation in Finland and found that the explanatory power of corporate and job segregation dropped from 23.6\% to 19.1\% between 1995 and 2004. Using data from the 1988-1992 Chinese Urban Household Survey, Ng \cite{ng2004economic} found that the total explanatory power of occupational segregation was 20\%-40\%, after controlling the influence of education, work experience, province, industry, occupation, and company ownership.


Up to now, most related studies rely on questionnaire survey \cite{landivar2013disparities} or experimental data \cite{2014The}, both of which are based on small samples. This work explores the relationship between occupational segregation and gender wage gap at occupational granularity by using large-scale samples. Our aim is to make practical suggestions to women about how to reduce the gender wage gap. The remainder of this paper is organized as follows. Section 2 presents the relationship between gender wage gaps and levels of occupational segregation in different occupations, as well as the overall severity of Chinese occupational segregation. Section 3 analyzes the severities of discrimination in different occupations by decomposing wages into those determined by human endowments and those caused by discrimination. Section 4 further decomposes gender wage gaps into intra- and inter-occupational differences. Finally, Section 5 briefs the conclusions along with some discussions.

\section{Gender Wage Gap and Occupational Segregation}
We use resume data of 10,318,484 job seekers crawled from various online recruitment websites from 2014 to 2015. The data includes basic personal information, educational experience, work experience and expected occupation and position. Among them, gender, age, previous wage and occupation from work experience are mainly utilized. A job seeker is accepted for further analysis if the information about gender, age, last year salary and occupation are all presented, and the age is at least 16 years old. After data screening, 3,266,272 job seekers are accepted, including 1,957,747 for males and 1,308,525 for females.

\begin{figure}[h]
    \centering
    \includegraphics[width=1\columnwidth]{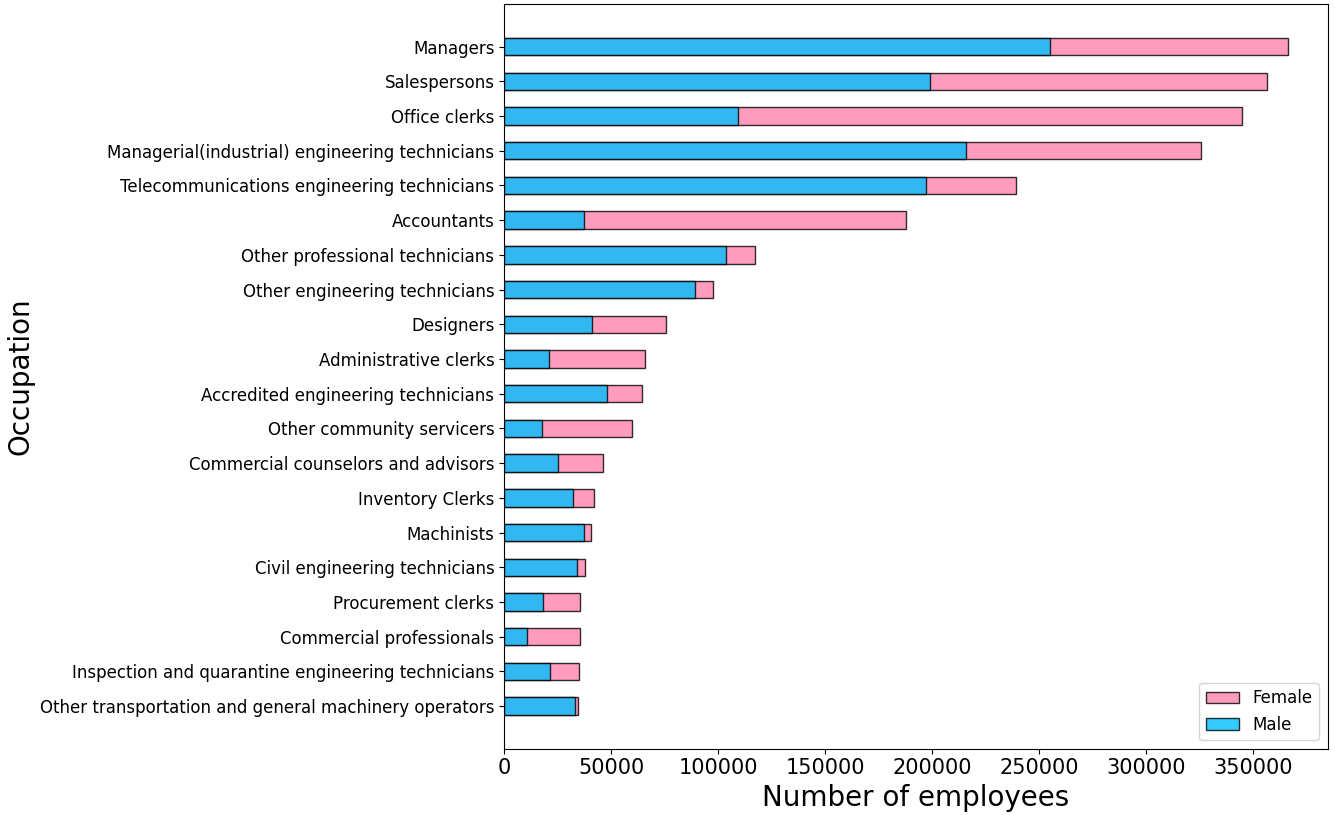}
    \caption{The number of employees of the top-20 occupations.}
    \label{fig0}
\end{figure}

In the following analysis, we mainly focus on the top-20 occupations with the largest number of employees. The total number of employees in those 20 occupations is 2,607,503, accounting for 79.83\% of the entire samples, as detailed in Fig. \ref{fig0}. 

\begin{figure}[h]
    \centering
    \includegraphics[width=0.6\columnwidth]{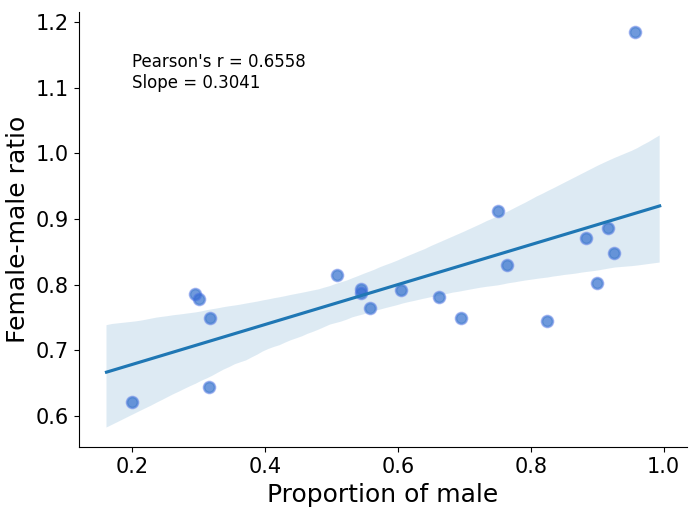}
    \caption{The relationship between proportions of male employees and female-male ratios in the top-20 occupations. The line is the linear fit of the scatter plot, and the blue shaded area is the 95\% confidence interval.}
    \label{fig1}
\end{figure}

\newpage
In Fig. \ref{fig1}, we present the relationship between gender wage gaps and severities of occupational segregation (measured by proportions of male employees) in the top-20 occupations. Though gender wage gaps exist (i.e., $r_{fm}<1$) in almost all occupations, it is observed that male-dominated occupations (such as engineering and professional technicians) have relatively smaller $r_{fm}$, while accountants, office clerks and other occupations that are generally perceived as more feminine have relatively larger $r_{fm}$. The result suggests that whether the occupation is dominated by men is significantly related to the gender wage gap (the Pearson correlation coefficient reaches 0.6558, with $p$-value$<$0.01 according to the Student's $t$-test). To some extent, the observed correlation indicates that although male-dominated occupations may have higher entry barriers for women, they exhibit narrower gender wage gap than female-dominated ones. 

\newpage
We calculate one of the most classic occupational segregation indices, the Ducan index \cite{duncan1955methodological}, which is a measure of what percentage of workers of another gender would have to change jobs in order to achieve an equal distribution of genders across occupations, when workers of one gender stayed in their current jobs. The Ducan index ranges from 0 to 1: if all occupations are completely dominated by one gender, it equals 1, while if proportions of male employees in all occupations are the same, it equals 0. The mathematical formula of the Ducan index is
\begin{equation}
    D={\frac {1}{2}} \sum_{j=1}^n |(F_j/F)-(M_j/M)|,
\end{equation}
where $F_j$ and $M_j$ are the number of female and male employees in occupation $j$ respectively, $F$ and $M$ are the number of all female and male employees, and $n$ is the number of considered occupations.

\begin{figure}[h]
    \centering
    \includegraphics[width=0.6\columnwidth]{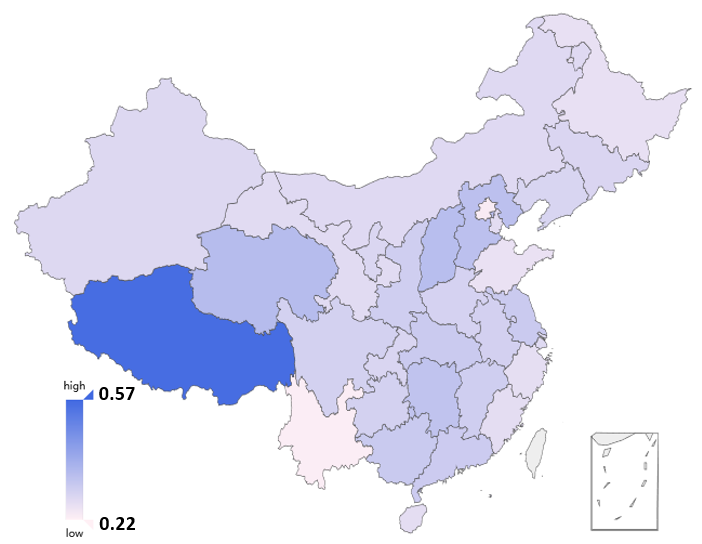}
    \caption{The Ducan indices for Chinese provinces.}
    \label{fig2}
\end{figure}

The Ducan index for the current data set is 0.398, which is remarkably lower than those of developed countries in the world (for example, the Ducan index in the United States has decreased from about 0.60 to 0.47 in the recent 60 years \cite{roos2018integrating}, and in Finland and Sweden, the Ducan index is still more than 0.40 in 2020 after a long decline \cite{mavrikiou2020impact}). At the same time, the Ducan index based on the 2010 Population Census of the People's Republic of China is 0.243, even lower than the resume data. Fig. \ref{fig2} shows the Ducan indices for different provinces with lighter color corresponding to less segregation. One can observe that, except for Tibet, severities of occupational segregation for different provinces are similar and low. As Tibet is an outlier, it is removed from the later statistics.

\begin{figure}[h]
    \centering
    \includegraphics[width=1\columnwidth]{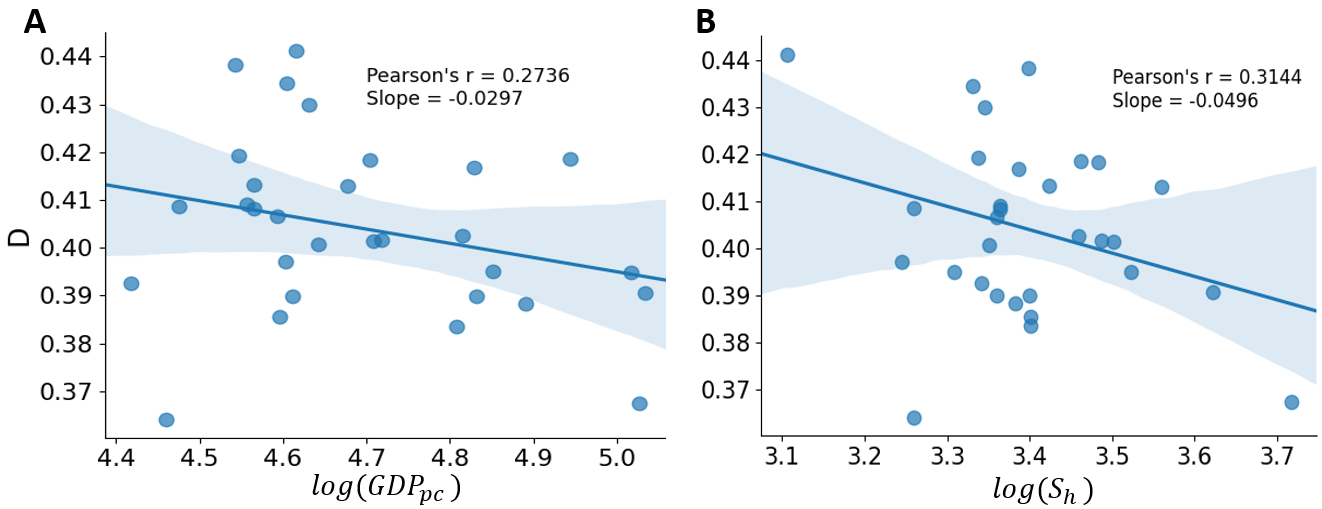}
    \caption{The relationship between $D$ and $GDP_{pc}$ (A), and the relationship between $D$ and $S_h$ (B) at the provincial level. The lines are the linear fits of the scatter plots, and the blue shaded areas are the 95\% confidence intervals.}
    \label{fig3}
\end{figure}

Fig. \ref{fig3} shows the relationships between the Ducan index and per capita GDP ($GDP_{pc}$) and the number of college students per $10^5$ people ($S_h$). As the majority of updatings in the resume data happened in 2015, the relevant statistical data at provincial level are taken from the China Statistical Yearbook 2016, which presents statistics of social and economic status in 2015. It can be seen that the Ducan index is significantly correlated with the level of economy and education ($p$-values are all less than 0.01 according to the Student's $t$-test). The above result suggests that the improvement of education and economy is helpful in reducing occupational segregation, however, the change of $D$ is not remarkable.

\section{Inter-occupational Gender Discrimination}
In order to better understand the impact of occupation-related discrimination on the gender wage gap, we implement Brown decomposition on the resume data \cite{brown1980equalizing}. Specifically, the gender wage gap is decomposed into differences within and between occupations, and gender occupational distribution is estimated from the perspective of job acquisition, thus separating out the inequality caused by gender-specific occupational entry barriers. In this paper, considering the distribution of individual occupations as an endogenous variable, we use the two-stage process and multiple choice model to estimate the entry probability of female (or male) in ``barrier-free" career choice. Finally, the Mincer wage equation is used to estimate the wage function of men and women in each occupation. Brown decomposition emphasizes the impact of occupational segregation on wage gaps and constructs a counterfactual framework where women face the same occupational structure as men. The mathematical formula of Brown decomposition is
\begin{equation}
\begin{aligned}
    \bar{S}_M-\bar{S}_F = &\underbrace{\sum\nolimits_j p_j^F (\bar{X}_j^M-\bar{X}_j^F) \beta_j^M}_{PD} + \underbrace{\sum\nolimits_j p_j^F \bar{X}_j^F(\bar{\beta}_j^M-\bar{\beta}_j^F)}_{WD} \\
    &+ \underbrace{\sum\nolimits_j \bar{S}_j^M(p_j^M-\tilde{p}_j^F)}_{QD} + \underbrace{\sum\nolimits_j \bar{S}_j^M(\tilde{p}_j^F - p_j^F)}_{OD},
\end{aligned}
\end{equation}
where $S$ represents the average annual wage, $X$ is the influencing factors of wage (i.e. the characteristic matrix), $p_j$ is the probability of an employee working in occupation $j$, $M$ and $F$ are short for male and female, and $\tilde{p}_j^F$ is the expected probability of $p_j^F$ if both occupational distributions of two genders are the same. PD and QD are the intra-occupational and inter-occupational gender differences of endowments, respectively. WD is intra-occupational gender wage discrimination, and OD accounts for the occupational segregation.

\begin{figure}[h]
    \centering
    \includegraphics[width=0.6\columnwidth]{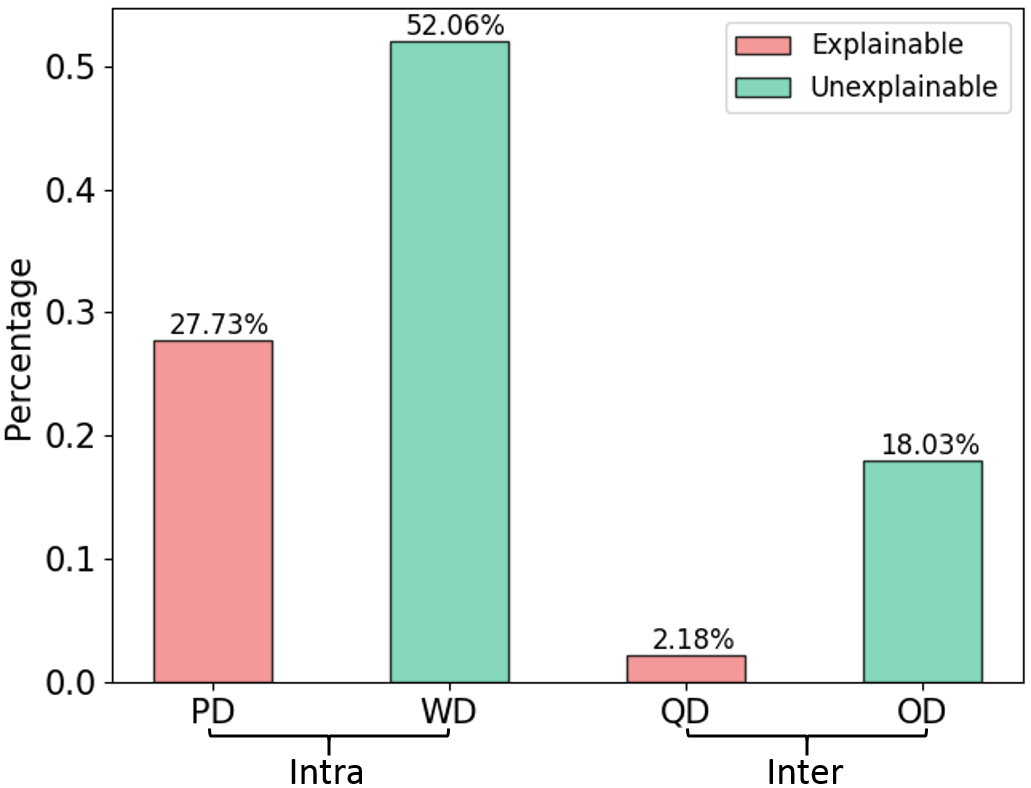}
    \caption{The result of Brown decomposition. PD and WD are intra-occupational differences, and the latter two are inter-occupational differences.}
    \label{fig5}
\end{figure}

The result of Brown decomposition is shown in Fig. \ref{fig5}, where WD accounts for the largest proportion (52.06\%), and QD accounts for the smallest proportion (2.18\%). It can be seen that the inter-occupational discrimination is low, which is in line with the fact that the Ducan index is also low. This indicates that occupations are not much segregated by gender, and thus women are relatively easy to jump to male-dominated occupations to avoid being trapped in occupations with very small $r_{fm}$.

\section{Intra-occupational Gender Discrimination}
According to the result of Brown decomposition, the intra-occupational gender discrimination is the main cause of the gender wage gap (79.79\%). We then apply the Blinder-Oaxaca (BO) decomposition for each occupation \cite{oaxaca1973male, blinder1973wage}, which decomposes wage gaps between gender groups into the explainable part caused by differences in individual characteristics, and the unexplainable part attributed to discrimination. Regression of the wage yields the following formula
\begin{equation}
\begin{aligned}
    &S_M = \beta_M X_M+\epsilon_M, \\
    &S_F = \beta_F X_F+\epsilon_F,
\end{aligned}
\end{equation}
where $\beta$ is the regression coefficient that captures the effects of human endowment on wage for male (or female), and $\epsilon$ is the error term. Then, the gap $1-r_{fm}$ can be expressed as
\begin{equation}
    1-r_{fm} \approx (\bar{X}_M-\bar{X}_F)\beta_M+(\bar{\beta}_M-\bar{\beta}_F)\bar{X}_F,
\end{equation}
where $(\bar{X}_M-\bar{X}_F)\beta_M$ represents the gap coming from the differences in individual characteristics assuming that there is no gender discrimination (explainable part), and $(\bar{\beta}_M-\bar{\beta}_F)\bar{X}_F$ is the gap resulted from discrimination (unexplainable part).

In the following analysis, we concentrate on job seekers whose gender, age, seniority, degree, school, last year salary, marital status, profession, industry, occupation, expected city, living city and home city are all known (in total there are 753,616 such job seekers). In addition, the wage is taken logarithm, and variables except salary and dummy variables are treated centrally before decomposition.

\begin{figure}[h]
    \centering
    \includegraphics[width=1\columnwidth]{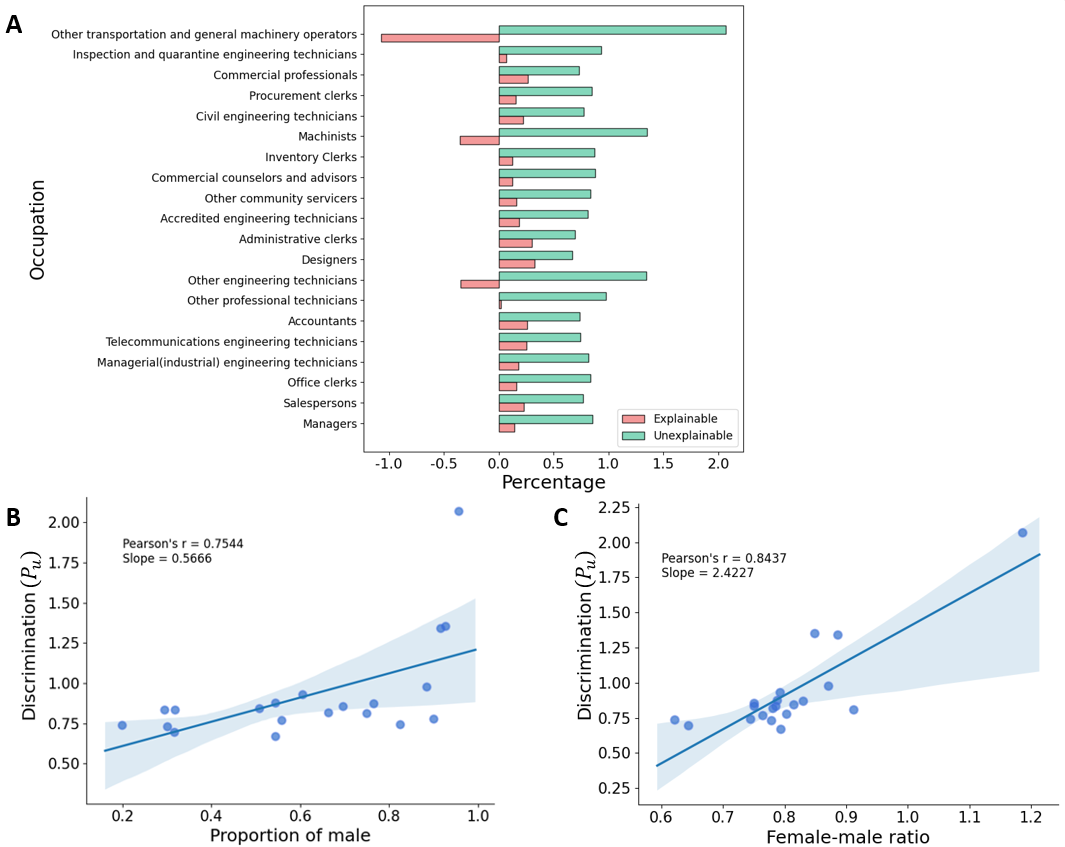}
    \caption{(A) The proportions of explainable ($P_e$) and unexplainable ($P_u$) parts in different occupations. (B) The relationship between the severities of occupational segregation and the proportions of the discrimination parts according to the BO decomposition. (C) The relationship between the gender wage gaps and the proportions of the discrimination parts according to the BO decomposition. The lines are the linear fits of the scatter plots, and the blue shaded areas are the 95\% confidence intervals.}
    \label{fig4}
\end{figure}

The proportions of the explainable and unexplainable parts are

\begin{equation}
\begin{aligned}
    P_e = \frac{(\bar{X}_M-\bar{X}_F)\beta_M}{(\bar{X}_M-\bar{X}_F)\beta_M+(\bar{\beta}_M-\bar{\beta}_F)\bar{X}_F}, \\
    P_u = \frac{(\bar{\beta}_M-\bar{\beta}_F)\bar{X}_F}{(\bar{X}_M-\bar{X}_F)\beta_M+(\bar{\beta}_M-\bar{\beta}_F)\bar{X}_F},
\end{aligned}
\end{equation}
Obviously, $P_e+P_u=1$ and $P_u$ is usually considered as discrimination. Overall speaking, according to the BO decomposition, $P_e$ is only 18.53\%, while $P_u = 81.47\%$.
Notice that, if $P_e<0$, for example $P_e=-0.1$, it means female employees should earn $0.1G$ more than male employees according to their individual characteristics, where $G$ denotes the gender wage gap. That is to say, the discrimination equals 1.1G, even larger than the observed gap. Fig. \ref{fig4}A shows the results of BO decomposition for the top-20 occupations, showing that the discrimination ($P_u$) of the gender wage gap is all relatively high, and even exceed 1 for a few occupations. Fig. \ref{fig4}B presents the relationship between the discrimination $P_u$ and the severities of occupational segregation, showing a strong and significant correlation ($p$-value$<$0.01 according to the Student's $t$-test). In addition, as shown in Fig. \ref{fig4}C, $P_u$ and $r_{fm}$ are also strongly correlated. In other words, although $r_{fm}$ in a male-dominated occupation may be smaller, the corresponding discrimination $P_u$ is very probably larger than a female-dominated occupation.

\newpage

\begin{figure}[h]
    \centering
    \includegraphics[width=0.6\columnwidth]{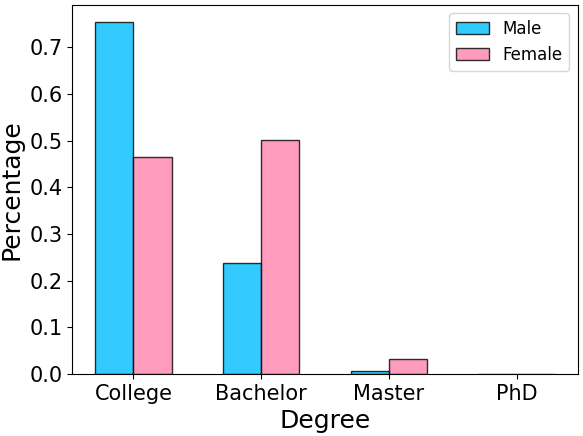}
    \caption{Distribution of the education experiences of male and female employees in ``other transportation and general machinery operators".}
    \label{fig6}
\end{figure}

In the above analysis, the occupation ``other transportation and general machinery operators" is an outlier. In this occupation, $r_{fm} = 1.19$, the proportion of men is 96\%, while $P_u>200\%$. Looking into the education experiences of employees in this occupation (see Fig. \ref{fig6}), we find that more than 75\% of men are junior college students, while more than half of women have a bachelor's degree or even above. This means that women’s high wage is achieved through a better educational background. While it is a good strategy to narrow the gender wage gap by jumping to male-dominated occupations, women also need to pay more efforts than men, and in those less-gap occupations, the discrimination may be even larger (see Fig. \ref{fig4}C).

\section{Conclusions and Discussion}
In this paper, using Chinese resume data from $\sim 3.3 \times 10^6$ online job seekers, we study the severity of occupational segregation in China and its impact on the gender wage gap. Our results show that the gender wage gap in male-dominated occupations is relatively small, while the occupational segregation is not serious (indicated by a smaller Ducan index than many other countries) and the inter-occupational discrimination is low (according to the Brown decomposition). Therefore, to join male-dominated occupations is a feasible way to narrow the gender wage gap. However, we also show that the occupations with smaller gender wage gaps usually suffer even larger gender discrimination. That is to say, in those occupations, female employees ought to earn much more than male employees according to their individual characteristics. As all results come from large-scale natural data, we believe the reported phenomena are statistically solid \cite{gao2019computational, zhou2021representative}.

Occupational segregation mainly comes from the innate physiological characteristics and physiques and the resulting differences in acquired skills of men and women. Men are physically superior to women, while women have innate advantages in communication and language. In terms of industry distribution, men are concentrated in manufacturing, construction and telecommunication sectors, while the proportion of women in services such as education and medical care is relatively high. Whether in male-dominated or female-dominated occupations, women’s wage and time spent are generally lower than men’s \cite{chuang2018gender, casado2020gender}. Studies have also shown that the occupational segregation may be due to women's reluctance to choose skilled occupations \cite{schneeweis2012girls}.

To reduce the occupational segregation, known studies mainly consider the demand side. For example, in terms of political management of enterprises, a gender quota system can be adopted to ensure a minimum proportion of female managers \cite{pande2012gender}. However, Shaikh \textit{et al.} \cite{2014Race} show that social policies don't seem to affect intra-group gender distributions and thus address inequality in wage distribution. Our results provide a solution from the supply side of the labor market. In contemporary China, if women want to narrow the gender wage gap, they can actively cultivate male characteristic skills, change the difference of individual endowment with men and jump to male-dominated occupations to obtain the same remuneration as men. But at the same time, women should also be prepared for more severe discrimination. Indeed, a lower gender wage gap is statistically associated with higher gender discrimination. Before the arrival to true general equality, the improvement of personal endowment seems to be the only credible and critical way to break the wage limit.

\bibliographystyle{elsarticle-num} 
\bibliography{cas-refs}


\end{document}